\documentclass{aa}
 
\usepackage{txfonts}
\usepackage{graphicx}
\usepackage{amssymb}
\usepackage{natbib}
\usepackage{array}
\usepackage{subfigure}
\usepackage{color}
\bibpunct{(}{)}{;}{a}{}{,}

\begin{document}

\title{An Improved Method for Estimating the Masses of Stars with Transiting Planets.}
\subtitle{}

\author{B.Enoch\inst{1}
 \and A.Collier Cameron\inst{1}
 \and N.R.Parley\inst{1}
 \and L.Hebb\inst{2}
}

\institute{School of Physics and Astronomy, University of St. Andrews, North Haugh, St Andrews, KY16 9SS
 \and Vanderbilt University, Department of Physics and Astronomy, Nashville, TN 37235
}

\authorrunning{B. Enoch et al.}

\date{Received / Accepted}

\abstract
{To determine the physical parameters of a transiting planet and its host star from photometric and spectroscopic analysis, it is essential to independently measure the stellar mass. This is often achieved by the use of evolutionary tracks and isochrones, but the mass result is only as reliable as the models used.}
{The recent paper by Torres et al (2009) showed that accurate values for stellar masses and radii could be obtained from a calibration using $T_{eff}$, log $g$ and $[Fe/H]$. We investigate whether a similarly good calibration can be obtained by substituting log $\rho$ - the fundamental parameter measured for the host star of a transiting planet - for log $g$, and apply this to star-exoplanet systems.}
{We perform a polynomial fit to stellar binary data provided in Torres et al (2009) to obtain the stellar mass and radius as functions of $T_{eff}$, log $\rho$ and $[Fe/H]$, with uncertainties on the fit produced from a Monte Carlo analysis. We apply the resulting equations to measurements for seventeen SuperWASP host stars, and also demonstrate the application of the calibration in a Markov Chain Monte Carlo analysis to obtain accurate system parameters where spectroscopic estimates of effective stellar temperature and metallicity are available.}
{We show that the calibration using log $\rho$ produces accurate values for the stellar masses and radii; we obtain masses and radii of the SuperWASP stars in good agreement with isochrone analysis results. We ascertain that the mass calibration is robust against uncertainties resulting from poor photometry, although a good estimate of stellar radius requires good-quality transit light curve to determine the duration of ingress and egress.}
{}


\maketitle 

\section{Introduction}

There are currently over 400 known exoplanets, of which more than 60 transit their host stars\footnotemark \footnotetext[1]{www.exoplanet.eu}. This important transiting subset are the only planets for which the orbital inclination, and hence precise stellar and planetary parameters, may be determined. The fundamental parameters found for the host star and transiting planet are stellar density (see below) and planetary surface gravity \citep{southworth04}. To convert these into values for the radii of both, it is necessary to find the stellar mass. This is often arrived at iteratively via deriving a stellar density from the lightcurve analysis and a stellar effective temperature from spectroscopy and using these with model evolutionary tracks and isochrones of appropriate metallicity to find a stellar mass and age \citep{sozzetti07}. Further photometric and spectroscopic analysis may then be performed to arrive at final values for the masses and radii of the star and planet, see e.g. \citet{hebb09}. The resulting values for masses and radii are therefore only as reliable as the evolutionary models used. A recent study by \citet{southworth09} highlighted the fact that discrepancies between different sets of evolutionary models represent the dominant source of systematic uncertainty in planetary parameters. For example, they find that the spread of mass values obtained for HD 209458 using different models is around $4\%$.

Here we develop a new one-step approach to determining the masses of exoplanet host stars from their effective temperatures, metallicities and photometric bulk densities. We base our method on the recent study by \citet{torres09} of the masses and radii of a large sample of well-characterised main-sequence stars belonging to non-interacting, eclipsing spectroscopic binaries. \citet{torres09} showed that accurate stellar masses and radii could be obtained using a calibration of stellar surface gravity, effective temperature and metallicity. They used a set of well-determined measurements of log $g$, $T_{eff}$, $[Fe/H]$, $M$ and $R$ from binary stars to obtain coefficients that allow mass and radius to be calculated directly for any normal star, without isochrone fitting. 

Recently, the use of log $\rho$ in place of log $g$ in the determination of star-planet system parameters has become widespread, see for example \citet{sozzetti07}, \citet{winn08b}, \citet{sozzetti09} and \citet{fernandez09}. Where high quality photometric data can be obtained of the transit event, the stellar parameters can be obtained more precisely using the stellar density value derived from the lightcurve than using the stellar surface gravity value from spectral analysis \citep{sozzetti07}. 
  
In Section 2 we review the methodology for determining exoplanet host-star densities from the transit geometry. We re-determine the mass and radius calibrations of \citet{torres09} using their data, and obtain comparably tight mass and radius calibrations using log $\rho$ in place of log $g$. In Section 3 we apply the method to the host stars of several transiting planets for which isochrone mass determinations have been published recently. In Section 4 we show how the method can be incorporated directly in a Markov-chain Monte Carlo (MCMC) analysis, to give the stellar mass as a derived parameter. 


\section{Analysis}

We used the tabulated data of \citet{torres09} to perform our calibration. Those data consist of 19 binary systems, i.e. 38 stars, for which the metallicity is known, after excluding systems that contain pre-Main Sequence stars. We fit a similar polynomial calibration, replacing log~$g$ with log~$\rho$, and apply it to star-exoplanet systems. The stellar density can be obtained directly from only photometric measurements via fitting of a transit event \citep{seager03}: the ratio of semi-major axis to stellar radius depends on the ratio of transit duration to orbital period via
\begin{equation}
\frac{a}{R_{\ast}} = \left[ \frac{(1+\sqrt{\Delta F})^2 - b^2 (1-sin^2 \frac{\pi T}{P})}{sin^2 \frac{\pi T}{P}} \right]^{1/2}
\end{equation}
\noindent where $\Delta F = (F - F_{transit})/F$, $b$ is the impact parameter, $T$ is the transit duration and $P$ is the orbital period.   

Combining this with Kepler's Third Law, 
\begin{equation}
P^2 = \frac{4 \pi^2 a^3}{G (M_{\ast} + M_p)}
\end{equation}
\noindent where $G$ is the gravitational constant, leads to an expression for the stellar density \citep{seager03}
\begin{equation}
\frac{\rho_{\ast}}{\rho_{\odot}} = \left[ \frac{4 \pi^2}{P^2 G} \right] \left[ \frac{(1 + \sqrt{\Delta F})^2 - b^2 (1 - sin^2 \frac{\pi T}{P} )}{sin^2 \frac{\pi T}{P}} \right]^{3/2}
\label{eq:dens}
\end{equation}
\noindent since $M_{\ast} >> M_p$ (this is strictly true for planets on circular orbits only). Thus the stellar density may be obtained directly from parameters measurable from a high-quality lightcurve: the duration and depth of transit, the impact parameter and the orbital period.

We used a Singular Value Decomposition (SVD) fit, weighted by error on the mass or radius measurements, to obtain coefficients on significant variables. The final fit gives coefficients for $X$ (= log$(T_{eff}) - 4.1$), $X^2$, log $\rho$, log $\rho^2$, log $\rho^3$ and $[Fe/H]$ for mass, and $X$, log $\rho$ and $[Fe/H]$ for radius (the secondary terms are insignificant for the radius fit). Thus the mass or radius may be computed by
\begin{equation}
\mbox{log} M = a_1 + a_2 X + a_3 X^2 + a_4 \mbox{log} \rho + a_5 \mbox{log} \rho^2 + a_6 \mbox{log} \rho^3 + a_7 [Fe/H]
\label{eq1}
\end{equation}
\begin{equation}
\mbox{log} R = b_1 + b_2 X + b_3 \mbox{log} \rho + b_4 [Fe/H]
\label{eq2}
\end{equation}

\begin{figure*}
  \subfigure[]
{
  \includegraphics[angle=90,width=3.6in,height=3in]{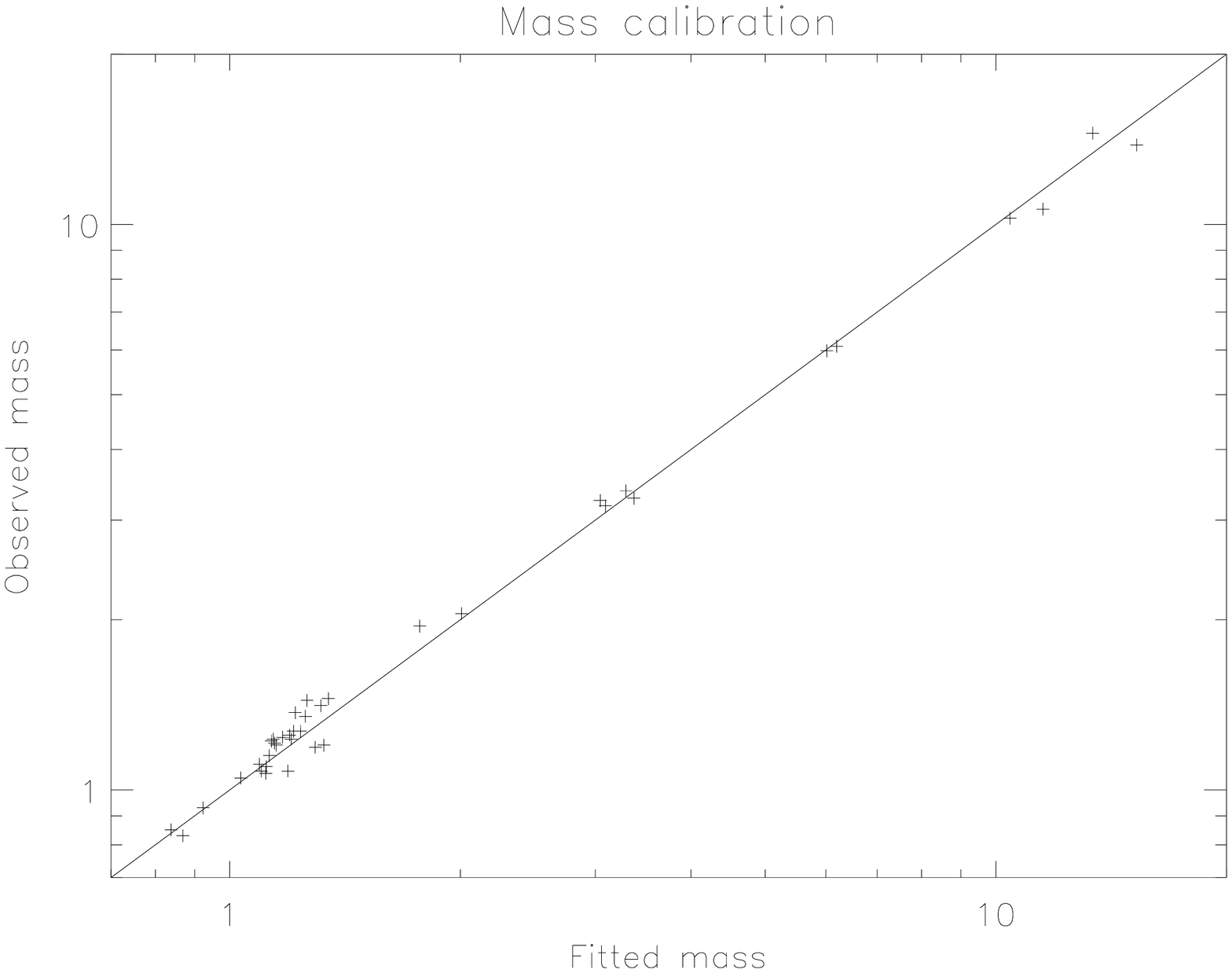}
}
\subfigure[]
{
  \includegraphics[angle=90,width=3.6in,height=3in]{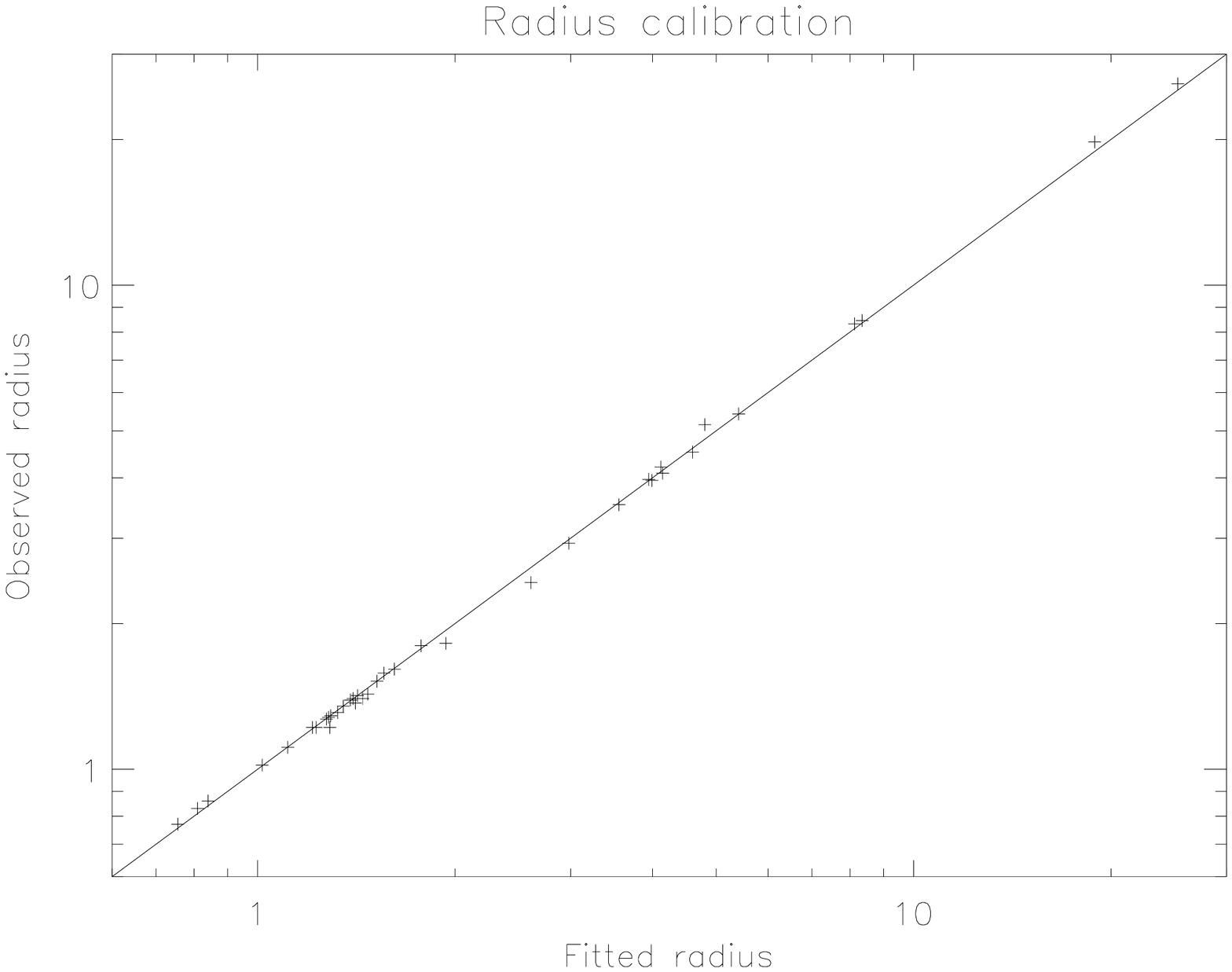}
}
\caption{Shows the scatter in fitted (calibrated) and measured (a) mass and (b) radius values for the 38 stars used in the calibration.}
\label{fig:massplot}
\end{figure*}

To obtain $1\sigma$ errors on the coefficients, we carried out a Monte Carlo analysis of 50,000 runs in which, for each run, each value for $T_{eff}$, log $\rho$, $[Fe/H]$ and $M$ or $R$ were perturbed randomly on a gaussian with standard deviation the $1\sigma$ error for that measurement. The set of 50,000 resulting coefficients were recorded, and the standard deviation of each coefficient within that set gave the error value for that coefficient. The fitted and error values for each of the mass and radius fits are given in Table \ref{table:coefs}. The resulting scatter in fitted less measured values is log $M$ = 0.023 and log $R$ = 0.009. Figure \ref{fig:massplot} shows calibrated versus measured mass and radius. 
 
\begin{center}
\begin{table}[!h]
\caption{Coefficients for Mass and Radius fits.}
\label{table:coefs}
\begin{tabular}{lrr}
\hline
 & Mass, $a_i$ & Radius, $b_i$ \\
\hline
const & $0.458\pm0.017$ & $0.150\pm0.002$ \\
$X$ & $1.430\pm0.019$ & $0.434\pm0.005$ \\
$X^2$ & $0.329\pm0.128$ & -  \\
log $\rho$ & $-0.042\pm0.021$ & $-0.381\pm0.002$ \\
log $\rho^2$ & $0.067\pm0.019$ & - \\
log $\rho^3$ & $0.010\pm0.004$ & - \\
$[Fe/H]$ & $0.044\pm0.019$ & $0.012\pm0.004$ \\
\hline
\end{tabular}
\end{table}
\end{center}


\begin{figure*}
  \subfigure[]
{
  \includegraphics[angle=90,width=3.6in,height=3in]{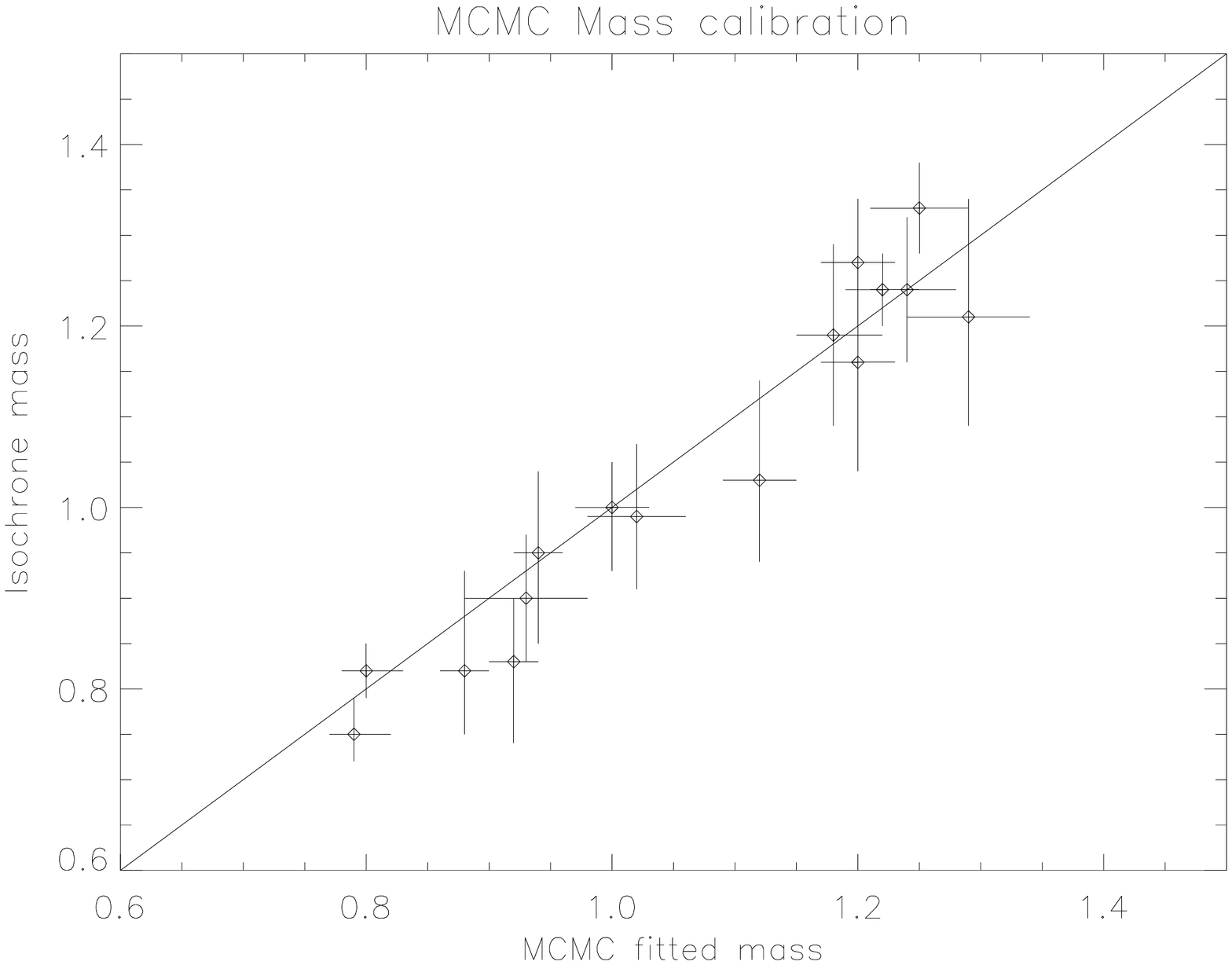}
}
\subfigure[]
{
  \includegraphics[angle=90,width=3.6in,height=3in]{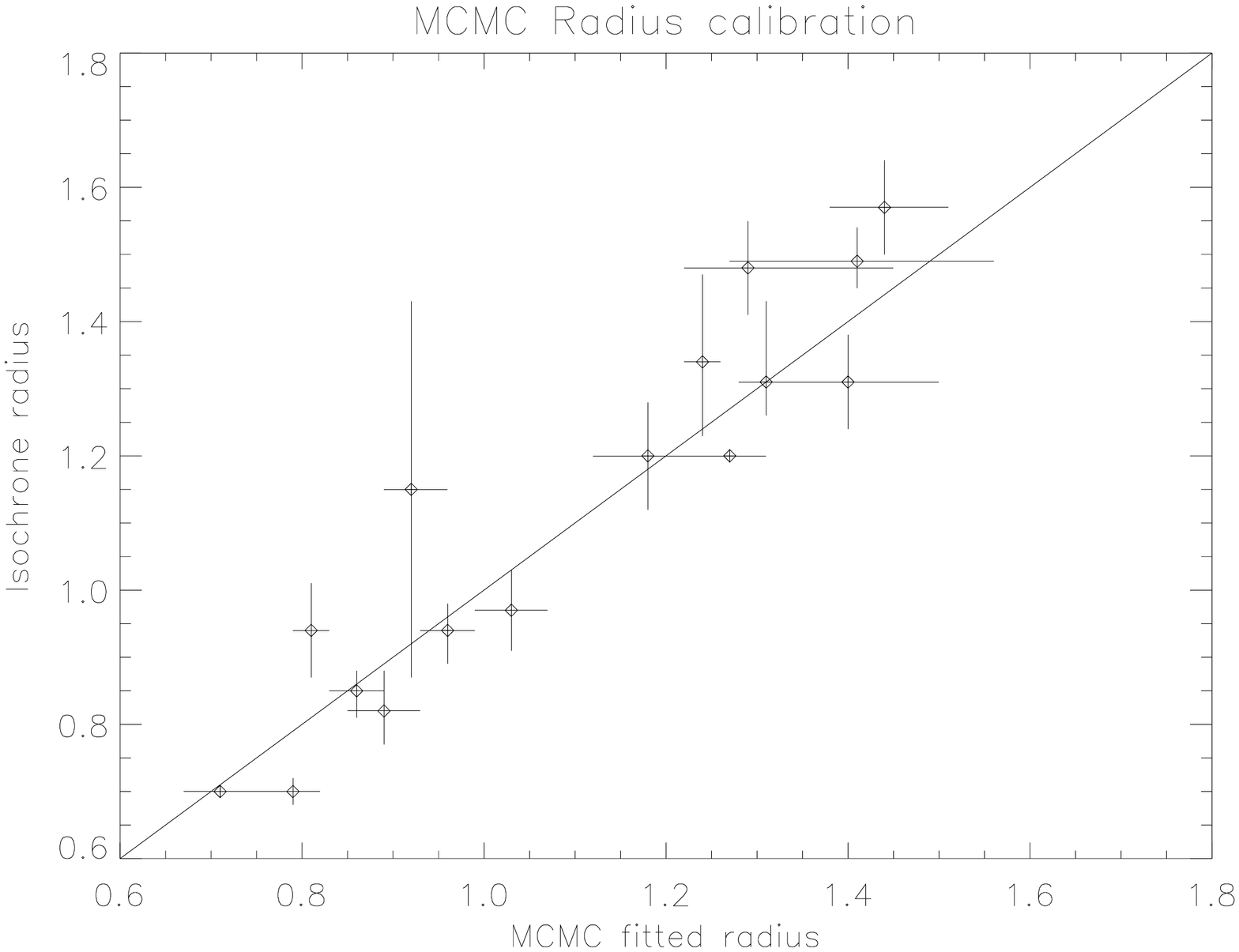}
}
\caption{Shows the scatter in isochrone (a) mass and (b) radius values versus values from MCMC analysis for 17 SuperWASP host stars.}
\label{fig:waspmassplot}
\end{figure*}




\section{Application to WASP host stars}

Seventeen WASP host stars have now been published. We use values for $T_{eff}$ and metallicity obtained from spectral analyses with the coefficients given in Table \ref{table:coefs} in Equations \ref{eq1} and \ref{eq2} to calculate the mass and radius of each, and compare these to the values obtained from isochrone interpolations. WASP-1 data was obtained from from \citet{cameron07b}, \citet{stempels07}, \citet{char07} and \citet{shporer07}, WASP-2 data from \citet{cameron07b} and \citet{char07}, WASP-3 data from \citet{pollacco08}, WASP-4 data from \citet{wilson08}, WASP-5 data from \citet{anderson08}, WASP-6 data from \citet{gillon09}, WASP-7 data from \citet{hellier09b}, WASP-10 data from \citet{johnson09} and \citet{christian09}, WASP-11 data from \citet{west09b} and \citet{bakos09}, WASP-12 data from \citet{hebb09}, WASP-13 data from \citet{skillen09}, WASP-14 data from \citet{joshi09}, WASP-15 data from \citet{west09}, WASP-16 data from \citet{lister09}, WASP-17 data from \citet{anderson09}, WASP-18 data from \citet{hellier09} and WASP-19 data from \citet{hebb10}. Table \ref{table:fits} shows that the agreement in results from the two methods is very good, demonstrating that computing stellar masses and radii from these coefficients is a valid, and simple, alternative to isochrone interpolation. 

The only values in Table \ref{table:fits} that do not quite agree, within errors, are the results for the radius of WASP-10. \citet{johnson09} find a radius of $0.70^{+0.01}_{-0.01} R_{\odot}$, whereas the calibration produces $R = 0.60^{+0.01}_{-0.01} R_{\odot}$. WASP-10 is an unusual host star, with a high density of $3.10 \rho_{\odot}$ \citep{johnson09} and a high level of activity \citep{smith09}. Such calibration discrepancies in low-mass, high-activity stars are discussed in Section 6 of \citet{torres09}. 

\begin{table*}
\begin{center}
\setlength{\extrarowheight}{5pt}
\caption{Comparison of mass and radius values obtained from isochrone fitting with those from the calibrated equations and from an MCMC analysis using those equations.}
\label{table:fits}
\begin{tabular}{lllrllllll}
\hline
 & & & & \multicolumn{2}{l}{Isochrone} & \multicolumn{2}{l}{Fitted} & \multicolumn{2}{l}{MCMC} \\
ID & $\rho_{\ast}$ & $T_{eff}$ & Metallicity & $M_{\ast}$ & $R_{\ast}$ & $M_{\ast}$ & $R_{\ast}$ & $M_{\ast}$ & $R_{\ast}$ \\
 & $\rho_{\odot}$ & K & & $M_{\odot}$ & $R_{\odot}$ & $M_{\odot}$ & $R_{\odot}$ & $M_{\odot}$ & $R_{\odot}$ \\
\hline
WASP-1  & $0.38^{+0.01}_{-0.01}$ & $6110\pm45$ & $0.26\pm0.03$ & $1.27^{+0.07}_{-0.07}$ & $1.49^{+0.05}_{-0.04}$ & $1.21^{+0.05}_{-0.07}$ & $1.50^{+0.02}_{-0.02}$ & $1.20^{+0.03}_{-0.03}$ & $1.41^{+0.15}_{-0.14}$  \\
WASP-2  & $1.51^{+0.16}_{-0.15}$ & $5150\pm80$ & $-0.08\pm0.08$ & $0.82^{+0.11}_{-0.07}$ & $0.82^{+0.06}_{-0.05}$ & $0.88^{+0.04}_{-0.06}$ & $0.82^{+0.03}_{-0.03}$ & $0.88^{+0.02}_{-0.02}$ & $0.89^{+0.04}_{-0.04}$ \\
WASP-3  & $0.55^{+0.15}_{-0.05}$ & $6400\pm100$ & $0.00\pm0.20$ & $1.24^{+0.08}_{-0.08}$ & $1.31^{+0.12}_{-0.05}$ & $1.21^{+0.05}_{-0.07}$ & $1.32^{+0.04}_{-0.09}$ & $1.22^{+0.04}_{-0.04}$ & $1.30^{+0.03}_{-0.04}$ \\
WASP-4  & $1.09^{+0.04}_{-0.09}$ & $5410\pm240$ & $0.00\pm0.20$ & $0.90^{+0.07}_{-0.07}$ & $1.15^{+0.28}_{-0.28}$ & $0.95^{+0.06}_{-0.08}$ & $0.95^{+0.03}_{-0.03}$ & $0.91^{+0.05}_{-0.05}$ & $0.88^{+0.02}_{-0.02}$ \\
WASP-5  & $1.08^{+0.22}_{-0.22}$ & $5700\pm150$ & $0.00\pm0.20$ & $0.99^{+0.08}_{-0.08}$ & $0.97^{+0.06}_{-0.06}$ & $1.01^{+0.05}_{-0.07}$ & $0.97^{+0.07}_{-0.09}$ & $1.01^{+0.04}_{-0.04}$ & $1.00^{+0.03}_{-0.02}$ \\
WASP-6  & $1.34^{+0.11}_{-0.10}$ & $5450\pm100$ & $-0.20\pm0.09$ & $0.83^{+0.07}_{-0.09}$ & $0.85^{+0.03}_{-0.04}$ & $0.93^{+0.04}_{-0.06}$ & $0.87^{+0.02}_{-0.03}$ & $0.93^{+0.02}_{-0.02}$ & $0.93^{+0.04}_{-0.04}$  \\
WASP-7  & $0.67^{+0.28}_{-0.11}$ & $6400\pm100$ & $0.00\pm0.10$ & $1.25^{+0.04}_{-0.08}$ & $1.23^{+0.09}_{-0.16}$ & $1.19^{+0.05}_{-0.07}$ & $1.23^{+0.07}_{-0.12}$ & $1.20^{+0.03}_{-0.03}$ & $1.21^{+0.04}_{-0.04}$ \\
WASP-10 & $3.10^{+0.09}_{-0.09}$ & $4675\pm100$ & $0.03\pm0.20$ & $0.75^{+0.04}_{-0.03}$ & $0.70^{+0.01}_{-0.01}$ & $0.80^{+0.05}_{-0.07}$ & $0.60^{+0.01}_{-0.01}$ & $0.79^{+0.02}_{-0.03}$ & $0.69^{+0.07}_{-0.03}$ \\
WASP-11 & $0.69^{+0.07}_{-0.11}$ & $4800\pm100$ & $0.00\pm0.20$ & $0.82^{+0.03}_{-0.03}$ & $0.75^{+0.02}_{-0.02}$ & $0.84^{+0.05}_{-0.06}$ & $0.76^{+0.02}_{-0.03}$ & $0.80^{+0.03}_{-0.02}$ & $0.79^{+0.03}_{-0.02}$ \\
WASP-12 & $0.35^{+0.03}_{-0.03}$ & $6300^{+200}_{-100}$ & $0.30^{+0.05}_{-0.15}$ & $1.33^{+0.05}_{-0.05}$ & $1.57^{+0.07}_{-0.07}$ & $1.27^{+0.06}_{-0.08}$ & $1.57^{+0.05}_{-0.07}$ & $1.28^{+0.05}_{-0.05}$ & $1.60^{+0.10}_{-0.10}$ \\
WASP-13 & $0.43^{+0.12}_{-0.10}$ & $5826\pm100$ & $0.00\pm0.20$ & $1.03^{+0.11}_{-0.09}$ & $1.34^{+0.13}_{-0.11}$ & $1.10^{+0.06}_{-0.08}$ & $1.40^{+0.15}_{-0.17}$ &  $1.11^{+0.03}_{-0.03}$ & $1.24^{+0.02}_{-0.02}$ \\
WASP-14 & $0.54^{+0.08}_{-0.06}$ & $6475\pm100$ & $0.00\pm0.20$ & $1.21^{+0.13}_{-0.12}$ & $1.31^{+0.07}_{-0.07}$ & $1.23^{+0.05}_{-0.07}$ & $1.34^{+0.05}_{-0.07}$  & $1.31^{+0.05}_{-0.10}$ & $1.49^{+0.07}_{-0.29}$ \\
WASP-15 & $0.37^{+0.04}_{-0.04}$ & $6300\pm100$ & $-0.17\pm0.11$ & $1.19^{+0.10}_{-0.10}$ & $1.48^{+0.07}_{-0.07}$ & $1.20^{+0.05}_{-0.07}$ & $1.52^{+0.05}_{-0.07}$  & $1.18^{+0.03}_{-0.03}$ & $1.37^{+0.10}_{-0.14}$ \\
WASP-16 & $1.21^{+0.13}_{-0.18}$ & $5700\pm150$ & $0.01\pm0.10$ & $1.00^{+0.05}_{-0.07}$ & $0.94^{+0.04}_{-0.05}$  & $1.01^{+0.05}_{-0.07}$ & $0.93^{+0.05}_{-0.05}$  & $1.00^{+0.03}_{-0.03}$ & $0.97^{+0.03}_{-0.03}$ \\
WASP-17 & $0.67^{+0.16}_{-0.13}$ & $6550\pm100$ & $-0.25\pm0.09$ & $1.16^{+0.12}_{-0.12}$ & $1.20^{+0.08}_{-0.08}$ & $1.19^{+0.05}_{-0.07}$ & $1.23^{+0.09}_{-0.11}$  & $1.19^{+0.03}_{-0.03}$ & $1.17^{+0.07}_{-0.06}$ \\
WASP-18 & $0.71^{+0.06}_{-0.10}$ & $6400\pm100$ & $0.00\pm0.09$ & $1.24^{+0.04}_{-0.04}$ & $1.20^{+0.06}_{-0.10}$  & $1.19^{+0.05}_{-0.06}$ & $1.12^{+0.06}_{-0.06}$ & $1.22^{+0.03}_{-0.03}$ & $1.27^{+0.04}_{-0.04}$ \\
WASP-19 & $1.13^{+0.12}_{-0.12}$ & $5500\pm100$ & $0.02\pm0.09$ & $0.95^{+0.09}_{-0.10}$ & $0.94^{+0.07}_{-0.07}$ & $0.97^{+0.04}_{-0.06}$ & $0.94^{+0.03}_{-0.04}$ & $0.93^{+0.02}_{-0.02}$ & $0.81^{+0.03}_{-0.02}$ \\
\hline
\end{tabular}
\end{center}
\end{table*}

\section{Application to Markov Chain Monte Carlo Analysis}

Having established that our modification of the Torres calibration yields stellar masses that agree well with those obtained from isochrone fitting, we now describe its implementation in our Markov-chain Monte Carlo (MCMC) parameter fitting code. This is followed by a discussion of a Main Sequence mass-radius constraint generally imposed on the parameter fitting, and objects for which this constraint is removed. We also discuss the effect on the final mass and radius values of uncertainty in eccentricity due to imperfect radial velocity and photometry. 

The MCMC analysis, described in \citet{cameron07} and \citet{pollacco08}, derives star-planet system parameters from simultaneous modelling of stellar lightcurve and radial velocity measurements. The host star`s radial velocity motion is parametrised by the radial velocity amplitude, the centre-of-mass velocity, the orbital eccentricity and the longitude of periastron, while the transit is modelled using the formulation of \citet{mandel02} incorporating the 4-coefficient limb-darkening model of \citet{claret00}. The form of the transit profile is determined by five parameters: the epoch of transit, orbital period, duration and depth of transit and the impact parameter of orbit. The best model is found using a constrained optimization of $\chi^2$ for both photometric and radial velocity data, combined with Bayesian priors on epoch, period, transit duration and depth, impact parameter, stellar mass, radial velocity semi-amplitude, eccentricity and the longitude of periastron. 

Here, we adapted this analysis to take the spectral $T_{eff}$ and metallicity as input values when modelling all available photometry and radial velocity observations of a host star. The MCMC code determines the stellar density value for the calculations from the observations at each step in the chain, now using Bayesian priors on the temperature and metallicity also, and uses the equations and coefficients above to calculate the stellar mass. These modifications make the MCMC analysis more robust since the stellar mass is now a derived quantity in the chain instead of a jump parameter constrained by a prior.

Where there is high-quality follow-up photometry of a transit event, the duration of ingress and egress, and hence the impact parameter, are tightly constrained. However, where such photometry is not available and thus the duration of ingress and egress cannot be accurately measured, an additional constraint is needed in the MCMC analysis. In such a case, the transit ingress and egress durations are overestimated in modelling the photometry since the sharp transitions to ingress and egress become blurred, allowing a shallower slope to be fitted. This leads to an underestimation of the impact parameter, $b$, from
\begin{equation}
b \approx 1 - \sqrt{\delta} \frac{T}{\tau}
\end{equation}
\noindent from \citet{winn09} where $\delta$ is the transit depth, $T$ is the total transit duration and $\tau$ is the partial, flat transit duration. This leads to an overestimation of the stellar radius, $R_{\ast}$ from
\begin{equation}
b = \frac{a \mbox{cos} i}{R_{\ast}}
\label{eq:imp}
\end{equation}
\noindent for a circular orbit, modified from \citet{winn09}, where $a$ is semi-major axis and $i$ is orbital inclination. To avoid this overestimation of stellar radius, the MCMC-fitted stellar radii are generally constrained to reasonably closely follow a Main Sequence relationship to the stellar mass, i.e. $R_{\ast} = M_{\ast}^{0.8}$ \citep{cox00}. To achieve this, a Bayesian prior is imposed on the stellar radius within the MCMC analysis such that a $\chi^2$ penalty is added to a chain step with a stellar radius different to that expected for the stellar mass value of that step, i.e.
\begin{equation}
\chi^2_{add} = \frac{(R_{step} - M_{\ast}^{0.8})^2}{\sigma_R^2} 
\end{equation}
\noindent where $R_{step}$ is the stellar radius value for the current step in the Monte Carlo chain, $M_{\ast}$ is the stellar mass of this chain step and $\sigma_R^2$ is the estimated uncertainty on the power-law estimate of the stellar radius. This $\chi^2$ penalty is added to the $\chi^2$ value for the chain step, giving the model a lower chance of being accepted.



However, WASP-1, 12 and 15 have this Main Sequence constraint relaxed due to being more evolved stars, each with an estimated age greater than their expected Main Sequence lifetime. WASP-1 and 12 are late F-type stars with an estimated main sequence lifetime of $\leq 1$Gyr but estimated ages of around 2 Gyr. The high-quality photometry of \citet{char07} and \citet{shporer07} (for WASP-1) and \citet{hebb09} (for WASP-12) constrains the impact parameter and stellar density satisfactorily, allowing a good estimate of stellar age, and at the same time removing the need for the additional MCMC constraint. WASP-15 is an F5 star of around 3.9 Gyr compared to a normal F5 age of 3.6 Gyr, also with high-quality photometry \citep{west09} which again constrains the impact parameter and stellar density. Restraining the radii of these stars to lower values with the Main Sequence constraint on leads to an overestimation of their density. In addition, we used the very high-quality photometry for WASP-10 given in \citet{johnson09}, and therefore had no need for the Main Sequence constraint in that case. For all other objects, without extremely high quality follow-up data and expected to be on the Main Sequence, the constraint was kept.  


With high precision photometry the precise shape of the transit would help constrain the eccentricity of the planetary orbit through the measured duration of transit. A planet moving on an eccentric orbit has its velocity modified by $(1+e\mbox{sin}\omega)/(\sqrt{1-e^2})$ compared to an identical planet on a circular orbit. Therefore for a planet on an eccentric orbit, the duration of transit, $T$, given by the length of the chord crossed in front of the star divided by the planet's velocity, is modified by $(\sqrt{1-e^2})/(1+e\mbox{sin}\omega)$ compared to the duration in a circular orbit. Without very high quality photometry, a source of uncertainty in the stellar density can arise from the fitting of the eccentricity of the planet`s orbit from the radial velocity measurements. With stellar density as an input, this leads to uncertainty in the results for stellar mass and radius using the calibrated equations and coefficients. This may especially be an issue for low mass planets which have a lower ratio of radial velocity amplitude to scatter than more massive planets, making an accurate determination of orbital eccentricity more difficult.  Using the simplified version of the equation for density, Equation 19 from \citet{seager03}, using $T\pi / P \ll 1$ so that sin$x \approx x$, and then including this factor gives
\begin{equation}
\frac{\rho_{\ast}}{\rho_{\odot}} \approx \frac{32}{G \pi}P \left[ \frac{\sqrt{\Delta F}}{ (T^2 - \tau^2) } \left( \frac{1+e\mbox{sin}\omega}{\sqrt{1-e^2} } \right) \right]^{3/2}
\end{equation}
\noindent where $\tau$ is the partial transit duration consisting of the `flat' part only.



As an example, WASP-13b is a planet of mass $0.46 M_J$ orbiting a G1 type star with a radial velocity semi-amplitude of $56 ms^{-1}$ \citep{skillen09}. The only transit photometry available is from SuperWASP and the 0.87m James Gregory Telescope in St Andrews, Scotland. Running the MCMC analysis as described above results in an eccentricity value of $e = 0.18\pm0.05$, with $\chi^2 = 15136$. However, holding the eccentricity fixed at 0 and repeating the analysis leads to an output with a very slightly higher $\chi^2$ value of 15140, so clearly the larger eccentricity value is not a secure result. The difference in resulting parameters are given in Table \ref{tab:wasp13}, showing that there is only a small effect of about six percent on the final stellar mass value despite a change of over 50\% in the stellar density. Therefore it may be concluded that the effect of the uncertainty in eccentricity on the final fitted mass value is small. The stellar radius is altered by around 18\%, due to the change in density value.


The results of the MCMC analysis on each of the 17 SuperWASP host stars are also presented in Table \ref{table:fits}, and are shown in Figure \ref{fig:waspmassplot}. Eccentricities are held fixed at 0 in all cases where this was done in the original papers WASP-3, 4, 5, 7, 11, 13, 15, 16 and 19). Almost all mass and radius values from the MCMC analysis agree with the isochrone values, within errors; the MCMC mass and radius ranges of WASP-6 and the radius range of WASP-19 do not quite overlap with the isochrone values, but all these values agree at the $2\sigma$ level.

\begin{table}
\setlength{\extrarowheight}{5pt}
\caption{Result of fixing $e = 0$ for WASP-13}
\label{tab:wasp13}
\begin{tabular}{ccccc}
\hline
$e$ & $b$ & $\rho_{\ast}$ & $M_{\ast}$ & $R_{\ast}$ \\ 
\hline
$0.18^{+0.05}_{0.05}$ & $0.15^{0.02}_{0.10}$ & $0.90^{0.14}_{0.11}$ & $1.05^{0.03}_{0.03}$ & $1.05^{0.05}_{0.05}$ \\
0.0 & $0.09^{0.10}_{0.07}$ & $0.58^{0.02}_{0.02}$ & $1.11^{0.03}_{0.03}$ & $1.24^{0.02}_{0.02}$ \\
\hline
\end{tabular}
\end{table}

\section{Summary}

We have presented a new calibration for stellar masses and radii based on stellar effective temperature, metallicity and stellar density. We have shown that the resulting equations provide a good fit to data for 38 stars from \citet{torres09}, and also to values for masses and radii of exoplanet host stars obtained from isochrone analyses. We have demonstrated that accurate stellar masses may be obtained for such exoplanet host stars via a Markov-chain Monte Carlo analysis of photometric and spectroscopic data, using spectroscopically determined temperatures and metallicities as input. 

Even where poor photometry yields an uncertain estimate of stellar density, the mass estimate from the calibration is encouragingly robust. However, the stellar radius depends strongly on the stellar density estimate which in turn requires good knowlege of the impact parameter. Thus in establishing planet radii there is no substitute for good quality photometry, though the Main Sequence prior can provide a useful additional constraint if the star can be shown via independent means to be unevolved. 

\begin{acknowledgements}
The WASP Consortium consists of astronomers primarily from the Queen's University Belfast, Keele, Leicester, The Open University, and St Andrews, the Isaac Newton Group (La Palma), the Instituto de Astrofisica de Canarias (Tenerife) and the South African Astronomical Observatory. The SuperWASP-N and WASP-S Cameras were constructed and operated with funds made available from Consortium Universities and the UK`s Science and Technology Facilities Council. 
\end{acknowledgements}

\bibliographystyle{aa}

\begin{thebibliography}{35}

\bibitem[{{Anderson} {et~al.}(2008){Anderson}, {Gillon}, {Hellier}, {Maxted},
  {Pepe}, {Queloz}, {Wilson}, {Collier Cameron}, {Smalley}, {Lister},
  {Bentley}, {Blecha}, {Christian}, {Enoch}, {Hebb}, {Horne}, {Irwin}, {Joshi},
  {Kane}, {Marmier}, {Mayor}, {Parley}, {Pollacco}, {Pont}, {Ryans},
  {S{\'e}gransan}, {Skillen}, {Street}, {Udry}, {West}, \&
  {Wheatley}}]{anderson08}
{Anderson}, D.~R., {Gillon}, M., {Hellier}, C., {et~al.} 2008, \mnras, 387, L4

\bibitem[{{Anderson} {et~al.}(2009){Anderson}, {Hellier}, {Gillon}, {Triaud},
  {Smalley}, {Hebb}, {Collier Cameron}, {Maxted}, {Queloz}, {West}, {Bentley},
  {Enoch}, {Horne}, {Lister}, {Mayor}, {Parley}, {Pepe}, {Pollacco},
  {S{\'e}gransan}, {Udry}, \& {Wilson}}]{anderson09}
{Anderson}, D.~R., {Hellier}, C., {Gillon}, M., {et~al.} 2009, ArXiv e-prints

\bibitem[{{Bakos} {et~al.}(2009){Bakos}, {P{\'a}l}, {Torres}, {Sip{\H o}cz},
  {Latham}, {Noyes}, {Kov{\'a}cs}, {Hartman}, {Esquerdo}, {Fischer}, {Johnson},
  {Marcy}, {Butler}, {Howard}, {Sasselov}, {Kov{\'a}cs}, {Stefanik},
  {L{\'a}z{\'a}r}, {Papp}, \& {S{\'a}ri}}]{bakos09}
{Bakos}, G.~{\'A}., {P{\'a}l}, A., {Torres}, G., {et~al.} 2009, \apj, 696, 1950

\bibitem[{{Cameron} {et~al.}(2007){Cameron}, {Bouchy}, {H{\'e}brard}, {Maxted},
  {Pollacco}, {Pont}, {Skillen}, {Smalley}, {Street}, {West}, {Wilson},
  {Aigrain}, {Christian}, {Clarkson}, {Enoch}, {Evans}, {Fitzsimmons},
  {Fleenor}, {Gillon}, {Haswell}, {Hebb}, {Hellier}, {Hodgkin}, {Horne},
  {Irwin}, {Kane}, {Keenan}, {Loeillet}, {Lister}, {Mayor}, {Moutou}, {Norton},
  {Osborne}, {Parley}, {Queloz}, {Ryans}, {Triaud}, {Udry}, \&
  {Wheatley}}]{cameron07b}
{Cameron}, A.~C., {Bouchy}, F., {H{\'e}brard}, G., {et~al.} 2007, \mnras, 375,
  951

\bibitem[{{Charbonneau} {et~al.}(2007){Charbonneau}, {Winn}, {Everett},
  {Latham}, {Holman}, {Esquerdo}, \& {O'Donovan}}]{char07}
{Charbonneau}, D., {Winn}, J.~N., {Everett}, M.~E., {et~al.} 2007, \apj, 658,
  1322

\bibitem[{{Christian} {et~al.}(2009){Christian}, {Gibson}, {Simpson}, {Street},
  {Skillen}, {Pollacco}, {Collier Cameron}, {Joshi}, {Keenan}, {Stempels},
  {Haswell}, {Horne}, {Anderson}, {Bentley}, {Bouchy}, {Clarkson}, {Enoch},
  {Hebb}, {H{\'e}brard}, {Hellier}, {Irwin}, {Kane}, {Lister}, {Loeillet},
  {Maxted}, {Mayor}, {McDonald}, {Moutou}, {Norton}, {Parley}, {Pont},
  {Queloz}, {Ryans}, {Smalley}, {Smith}, {Todd}, {Udry}, {West}, {Wheatley}, \&
  {Wilson}}]{christian09}
{Christian}, D.~J., {Gibson}, N.~P., {Simpson}, E.~K., {et~al.} 2009, \mnras,
  392, 1585

\bibitem[{{Claret}(2000)}]{claret00}
{Claret}, A. 2000, VizieR Online Data Catalog, 336, 31081

\bibitem[{{Collier Cameron} {et~al.}(2007){Collier Cameron}, {Wilson}, {West},
  {Hebb}, {Wang}, {Aigrain}, {Bouchy}, {Christian}, {Clarkson}, {Enoch},
  {Esposito}, {Guenther}, {Haswell}, {H{\'e}brard}, {Hellier}, {Horne},
  {Irwin}, {Kane}, {Loeillet}, {Lister}, {Maxted}, {Mayor}, {Moutou}, {Parley},
  {Pollacco}, {Pont}, {Queloz}, {Ryans}, {Skillen}, {Street}, {Udry}, \&
  {Wheatley}}]{cameron07}
{Collier Cameron}, A., {Wilson}, D.~M., {West}, R.~G., {et~al.} 2007, \mnras,
  380, 1230

\bibitem[{Cox(2000)}]{cox00}
Cox, A.~N., ed. 2000, {Allen's Astrophysical Quantities}, 4th edn. (Springer,
  Berlin)

\bibitem[{{Fernandez} {et~al.}(2009){Fernandez}, {Holman}, {Winn}, {Torres},
  {Shporer}, {Mazeh}, {Esquerdo}, \& {Everett}}]{fernandez09}
{Fernandez}, J.~M., {Holman}, M.~J., {Winn}, J.~N., {et~al.} 2009, \aj, 137,
  4911

\bibitem[{{Gillon} {et~al.}(2009){Gillon}, {Anderson}, {Triaud}, {Hellier},
  {Maxted}, {Pollaco}, {Queloz}, {Smalley}, {West}, {Wilson}, {Bentley},
  {Collier Cameron}, {Enoch}, {Hebb}, {Horne}, {Irwin}, {Joshi}, {Lister},
  {Mayor}, {Pepe}, {Parley}, {Segransan}, {Udry}, \& {Wheatley}}]{gillon09}
{Gillon}, M., {Anderson}, D.~R., {Triaud}, A.~H.~M.~J., {et~al.} 2009, \aap,
  501, 785

\bibitem[{{Hebb} {et~al.}(2009){Hebb}, {Collier-Cameron}, {Loeillet},
  {Pollacco}, {H{\'e}brard}, {Street}, {Bouchy}, {Stempels}, {Moutou},
  {Simpson}, {Udry}, {Joshi}, {West}, {Skillen}, {Wilson}, {McDonald},
  {Gibson}, {Aigrain}, {Anderson}, {Benn}, {Christian}, {Enoch}, {Haswell},
  {Hellier}, {Horne}, {Irwin}, {Lister}, {Maxted}, {Mayor}, {Norton}, {Parley},
  {Pont}, {Queloz}, {Smalley}, \& {Wheatley}}]{hebb09}
{Hebb}, L., {Collier-Cameron}, A., {Loeillet}, B., {et~al.} 2009, \apj, 693,
  1920

\bibitem[{{Hebb} {et~al.}(2010){Hebb}, {Collier-Cameron}, {Triaud}, {Lister},
  {Smalley}, {Maxted}, {Hellier}, {Anderson}, {Pollacco}, {Gillon}, {Queloz},
  {West}, {Bentley}, {Enoch}, {Haswell}, {Horne}, {Mayor}, {Pepe}, {Segransan},
  {Skillen}, {Udry}, \& {Wheatley}}]{hebb10}
{Hebb}, L., {Collier-Cameron}, A., {Triaud}, A.~H.~M.~J., {et~al.} 2010, \apj,
  708, 224

\bibitem[{{Hellier} {et~al.}(2009{\natexlab{a}}){Hellier}, {Anderson},
  {Cameron}, {Gillon}, {Hebb}, {Maxted}, {Queloz}, {Smalley}, {Triaud}, {West},
  {Wilson}, {Bentley}, {Enoch}, {Horne}, {Irwin}, {Lister}, {Mayor}, {Parley},
  {Pepe}, {Pollacco}, {Segransan}, {Udry}, \& {Wheatley}}]{hellier09}
{Hellier}, C., {Anderson}, D.~R., {Cameron}, A.~C., {et~al.}
  2009{\natexlab{a}}, \nat, 460, 1098

\bibitem[{{Hellier} {et~al.}(2009{\natexlab{b}}){Hellier}, {Anderson},
  {Gillon}, {Lister}, {Maxted}, {Queloz}, {Smalley}, {Triaud}, {West},
  {Wilson}, {Alsubai}, {Bentley}, {Cameron}, {Hebb}, {Horne}, {Irwin}, {Kane},
  {Mayor}, {Pepe}, {Pollacco}, {Skillen}, {Udry}, {Wheatley}, {Christian},
  {Enoch}, {Haswell}, {Joshi}, {Norton}, {Parley}, {Ryans}, {Street}, \&
  {Todd}}]{hellier09b}
{Hellier}, C., {Anderson}, D.~R., {Gillon}, M., {et~al.} 2009{\natexlab{b}},
  \apjl, 690, L89

\bibitem[{{Johnson} {et~al.}(2009){Johnson}, {Winn}, {Cabrera}, \&
  {Carter}}]{johnson09}
{Johnson}, J.~A., {Winn}, J.~N., {Cabrera}, N.~E., \& {Carter}, J.~A. 2009,
  \apjl, 692, L100

\bibitem[{{Joshi} {et~al.}(2009){Joshi}, {Pollacco}, {Cameron}, {Skillen},
  {Simpson}, {Steele}, {Street}, {Stempels}, {Christian}, {Hebb}, {Bouchy},
  {Gibson}, {H{\'e}brard}, {Keenan}, {Loeillet}, {Meaburn}, {Moutou},
  {Smalley}, {Todd}, {West}, {Anderson}, {Bentley}, {Enoch}, {Haswell},
  {Hellier}, {Horne}, {Irwin}, {Lister}, {McDonald}, {Maxted}, {Mayor},
  {Norton}, {Parley}, {Perrier}, {Pont}, {Queloz}, {Ryans}, {Smith}, {Udry},
  {Wheatley}, \& {Wilson}}]{joshi09}
{Joshi}, Y.~C., {Pollacco}, D., {Cameron}, A.~C., {et~al.} 2009, \mnras, 392,
  1532

\bibitem[{{Lister} {et~al.}(2009){Lister}, {Anderson}, {Gillon}, {Hebb},
  {Smalley}, {Triaud}, {Collier Cameron}, {Wilson}, {West}, {Bentley},
  {Christian}, {Enoch}, {Haswell}, {Hellier}, {Horne}, {Irwin}, {Joshi},
  {Kane}, {Mayor}, {Maxted}, {Norton}, {Parley}, {Pepe}, {Pollacco}, {Queloz},
  {Ryans}, {Segransan}, {Skillen}, {Street}, {Todd}, {Udry}, \&
  {Wheatley}}]{lister09}
{Lister}, T.~A., {Anderson}, D.~R., {Gillon}, M., {et~al.} 2009, \apj, 703, 752

\bibitem[{{Mandel} \& {Agol}(2002)}]{mandel02}
{Mandel}, K. \& {Agol}, E. 2002, \apjl, 580, L171

\bibitem[{{Pollacco} {et~al.}(2008){Pollacco}, {Skillen}, {Collier Cameron},
  {Loeillet}, {Stempels}, {Bouchy}, {Gibson}, {Hebb}, {H{\'e}brard}, {Joshi},
  {McDonald}, {Smalley}, {Smith}, {Street}, {Udry}, {West}, {Wilson},
  {Wheatley}, {Aigrain}, {Alsubai}, {Benn}, {Bruce}, {Christian}, {Clarkson},
  {Enoch}, {Evans}, {Fitzsimmons}, {Haswell}, {Hellier}, {Hickey}, {Hodgkin},
  {Horne}, {Hrudkov{\'a}}, {Irwin}, {Kane}, {Keenan}, {Lister}, {Maxted},
  {Mayor}, {Moutou}, {Norton}, {Osborne}, {Parley}, {Pont}, {Queloz}, {Ryans},
  \& {Simpson}}]{pollacco08}
{Pollacco}, D., {Skillen}, I., {Collier Cameron}, A., {et~al.} 2008, \mnras,
  385, 1576

\bibitem[{{Seager} \& {Mall{\'e}n-Ornelas}(2003)}]{seager03}
{Seager}, S. \& {Mall{\'e}n-Ornelas}, G. 2003, \apj, 585, 1038

\bibitem[{{Shporer} {et~al.}(2007){Shporer}, {Tamuz}, {Zucker}, \&
  {Mazeh}}]{shporer07}
{Shporer}, A., {Tamuz}, O., {Zucker}, S., \& {Mazeh}, T. 2007, \mnras, 376,
  1296

\bibitem[{{Skillen} {et~al.}(2009){Skillen}, {Pollacco}, {Collier Cameron},
  {Hebb}, {Simpson}, {Bouchy}, {Christian}, {Gibson}, {H{\'e}brard}, {Joshi},
  {Loeillet}, {Smalley}, {Stempels}, {Street}, {Udry}, {West}, {Anderson},
  {Barros}, {Enoch}, {Haswell}, {Hellier}, {Horne}, {Irwin}, {Keenan},
  {Lister}, {Maxted}, {Mayor}, {Moutou}, {Norton}, {Parley}, {Queloz}, {Ryans},
  {Todd}, {Wheatley}, \& {Wilson}}]{skillen09}
{Skillen}, I., {Pollacco}, D., {Collier Cameron}, A., {et~al.} 2009, \aap, 502,
  391

\bibitem[{{Smith} {et~al.}(2009){Smith}, {Hebb}, {Collier Cameron}, {Anderson},
  {Lister}, {Hellier}, {Pollacco}, {Queloz}, {Skillen}, \& {West}}]{smith09}
{Smith}, A.~M.~S., {Hebb}, L., {Collier Cameron}, A., {et~al.} 2009, \mnras,
  398, 1827

\bibitem[{{Southworth}(2009)}]{southworth09}
{Southworth}, J. 2009, \mnras, 394, 272

\bibitem[{{Southworth} {et~al.}(2004){Southworth}, {Zucker}, {Maxted}, \&
  {Smalley}}]{southworth04}
{Southworth}, J., {Zucker}, S., {Maxted}, P.~F.~L., \& {Smalley}, B. 2004,
  \mnras, 355, 986

\bibitem[{{Sozzetti} {et~al.}(2007){Sozzetti}, {Torres}, {Charbonneau},
  {Latham}, {Holman}, {Winn}, {Laird}, \& {O'Donovan}}]{sozzetti07}
{Sozzetti}, A., {Torres}, G., {Charbonneau}, D., {et~al.} 2007, \apj, 664, 1190

\bibitem[{{Sozzetti} {et~al.}(2009){Sozzetti}, {Torres}, {Charbonneau}, {Winn},
  {Korzennik}, {Holman}, {Latham}, {Laird}, {Fernandez}, {O'Donovan},
  {Mandushev}, {Dunham}, {Everett}, {Esquerdo}, {Rabus}, {Belmonte}, {Deeg},
  {Brown}, {Hidas}, \& {Baliber}}]{sozzetti09}
{Sozzetti}, A., {Torres}, G., {Charbonneau}, D., {et~al.} 2009, \apj, 691, 1145

\bibitem[{{Stempels} {et~al.}(2007){Stempels}, {Collier Cameron}, {Hebb},
  {Smalley}, \& {Frandsen}}]{stempels07}
{Stempels}, H.~C., {Collier Cameron}, A., {Hebb}, L., {Smalley}, B., \&
  {Frandsen}, S. 2007, \mnras, 379, 773

\bibitem[{{Torres} {et~al.}(2009){Torres}, {Andersen}, \&
  {Gim{\'e}nez}}]{torres09}
{Torres}, G., {Andersen}, J., \& {Gim{\'e}nez}, A. 2009, \aapr, 13

\bibitem[{{West} {et~al.}(2009{\natexlab{a}}){West}, {Anderson}, {Gillon},
  {Hebb}, {Hellier}, {Maxted}, {Queloz}, {Smalley}, {Triaud}, {Wilson},
  {Bentley}, {Collier Cameron}, {Enoch}, {Horne}, {Irwin}, {Lister}, {Mayor},
  {Parley}, {Pepe}, {Pollacco}, {Segransan}, {Spano}, {Udry}, \&
  {Wheatley}}]{west09}
{West}, R.~G., {Anderson}, D.~R., {Gillon}, M., {et~al.} 2009{\natexlab{a}},
  \aj, 137, 4834

\bibitem[{{West} {et~al.}(2009{\natexlab{b}}){West}, {Collier Cameron}, {Hebb},
  {Joshi}, {Pollacco}, {Simpson}, {Skillen}, {Stempels}, {Wheatley}, {Wilson},
  {Anderson}, {Bentley}, {Bouchy}, {Christian}, {Enoch}, {Gibson},
  {H{\'e}brard}, {Hellier}, {Loeillet}, {Mayor}, {Maxted}, {McDonald},
  {Moutou}, {Pont}, {Queloz}, {Smith}, {Smalley}, {Street}, \&
  {Udry}}]{west09b}
{West}, R.~G., {Collier Cameron}, A., {Hebb}, L., {et~al.} 2009{\natexlab{b}},
  \aap, 502, 395

\bibitem[{{Wilson} {et~al.}(2008){Wilson}, {Gillon}, {Hellier}, {Maxted},
  {Pepe}, {Queloz}, {Anderson}, {Collier Cameron}, {Smalley}, {Lister},
  {Bentley}, {Blecha}, {Christian}, {Enoch}, {Haswell}, {Hebb}, {Horne},
  {Irwin}, {Joshi}, {Kane}, {Marmier}, {Mayor}, {Parley}, {Pollacco}, {Pont},
  {Ryans}, {Segransan}, {Skillen}, {Street}, {Udry}, {West}, \&
  {Wheatley}}]{wilson08}
{Wilson}, D.~M., {Gillon}, M., {Hellier}, C., {et~al.} 2008, \apjl, 675, L113

\bibitem[{{Winn}(2009)}]{winn09}
{Winn}, J.~N. 2009, in IAU Symposium, Vol. 253, IAU Symposium, 99--109

\bibitem[{{Winn} {et~al.}(2008){Winn}, {Holman}, {Torres}, {McCullough},
  {Johns-Krull}, {Latham}, {Shporer}, {Mazeh}, {Garcia-Melendo}, {Foote},
  {Esquerdo}, \& {Everett}}]{winn08b}
{Winn}, J.~N., {Holman}, M.~J., {Torres}, G., {et~al.} 2008, \apj, 683, 1076

\end{thebibliography}

\end{document}